\documentclass[%
reprint,
superscriptaddress,
amsmath,amssymb,
aps,
pra,
]{revtex4-2}

\usepackage{graphicx}
\usepackage{dcolumn}
\usepackage{bm}
\usepackage{float}

\begin{document}
	
\preprint{APS/123-QED}
	
\title{Plasmonic band and defect mode of one dimensional graphene lattice}
\author{Yun-Cheng Zhou}
	
\affiliation{School of Physics, Huazhong University of Science and Technology, Wuhan 430074, China
	}
\author{Xiaodong Zeng}%
\affiliation{Department of Physics, Shanghai University, Shanghai 200444, China}
\author{Rafi Ud Din}
\affiliation{School of Physics, Huazhong University of Science and Technology, Wuhan 430074, China
	}
	\author{Guo-Qin Ge}
	\email{gqge@hust.edu.cn}
	\affiliation{School of Physics, Huazhong University of Science and Technology, Wuhan 430074, China
	}
	
	\author{Muhammad Suhail Zubairy}
	\affiliation{Institute for Quantum Science and Engineering (IQSE) and Department of Physics and Astronomy, Texas A\&M University, College Station, Texas 77843-4242, USA
	}

	
	\date{\today}
	
	\begin{abstract}
		Photonic crystals based on graphene plasmons (GPs) are highly tunable and can accurately control photonic transmission at the nanoscales. In this work, the transfer matrix method (TMM) is introduced to study graphene plasmonic crystal (GPC) with periodic surface conductivity in the case of normal incidence.
		The introduction of TMM after considering the abnormal phase scattering of the abrupt interface gives an idea to accurately manipulate plasmonic crystal structures, and can reduce the calculation workload to a certain extent. The effectiveness of the proposed method is verified with the plane wave expansion method in our model. 
		Furthermore, we study the defect mode and the plasmonic Tamm state in GPC by the transfer matrix method. 
	\end{abstract}
	
	\pacs{Valid PACS appear here}
	\maketitle
	
	
	\section{\label{sec:level1}INTRODUCTION}
	Graphene, a honeycomb lattice of carbon, has achieved tremendous breakthroughs in various fields ranging from condensed matter physics and materials science to physical electronics, mechanics and thermal processes. Its highly carrier mobility, intense surface-plasmon mode confinement and excellent electrical tunability lead graphene to an extremely capable plasmonic material~ \cite{dias2016introduction,basov2014colloquium}. It supports highly confined plasmon polaritons, i.e., graphene plasmons (GPs) with very low dissipative losses~\cite{PhysRevLett.58.2059,PhysRevLett.58.2486,meade1995photonic,naik2013alternative}. Being a two-dimensional membrane, graphene is readily for stacking. For example, a photonic crystal can be constructed by stacking several sheets of graphene separated by dielectric layers~\cite{Bludov2016,Fu2016,Sayed2020}. Photonic crystals are synthetic structures, essentially designed and fabricated to emulate advanced physical concepts.~\cite{Sunku2018}. However, conventional photonic crystals cannot be readily turned on/off, and have typically limited tunability. Furthermore, it is possible to carry out a graphene plasmonic crystal (GPC), a photonic crystal composed of plasmonic materials~\cite{RN39,PhysRevLett.102.146807,Wang:18}. It can induce an energy band structure analogous to superlattices and photonic crystals~\cite{Wang2019}.
	
	GPCs based on monolayer graphenes have also been recently proposed and developed. In these structures, a periodic Fermi energy landscape can be imposed on graphene by placing a designed metagate at a distance of few nanometers from graphene. This GPC was shown to produce complete propagation band gaps for GPs under highly suppressed scattering~\cite{Jung2018}. A spacer can also be inserted between the metagate and graphene. The different spacer thicknesses induce different Fermi energies under a back-gate bias voltage~\cite{Wang2019}. The general Bloch-Floquet theorem in the periodically modulated equilibrium electron density of doped graphene induces energy gaps for GPs, which also form a GPC~\cite{RN40}. Notably, such structures lead to the scattering of plasmons, which should be deeply understood for controlling light propagation and confinement in integrated optical circuits. In recent years, many interesting studies have been conducted to study GPs in synthetic structures of graphene. Using transmission matrix method, a GPC was proposed where the mode transmission profile was modified by changing the chemical potential of the graphene~\cite{Fu2016}. Very recently, the scattering of one-dimensional quasistatic plasmons in periodic graphene structures with junctions of three different types was demonstrated from the numerical solutions of Maxwell equations~\cite{Semenenko2020}. It was usually assumed that the phase picked up by the GPs scatting at abrupt edges is $-\pi$, however, Nikitin et al. found that the reflection phase is about $-3\pi/4$~\cite{PhysRevB.90.041407}. This non-trivial phase comes from the complex excitation of highly evanescent modes near the edge, which satisfies the continuity of electric and magnetic fields. An analytical description of the abnormal phase was found in ~\cite{Rejaei_2015}. The scattering of GPs at the interface between two semi-infinite graphene sheets with different doping levels and/or different underlying dielectric substrates was also conducted and was found that there is almost no free radiation in the scattering process~\cite{PhysRevB.97.035434}. For normal incidence of GPs along a one-dimensional junction in a graphene sheet, two types of reflection were noticed, i.e., strong and weak reflection caused by resonance~\cite{Jiang:18}.
	
	Patterned GPs structures have been realized experimentally and found applications as phase modulators~\cite{RN32}, which requires only 350nm device length to adjust the phase in situ between 0 and $2\pi$. Xiong et al. also experimentally demonstrated a tunable two-dimensional photonic crystal for surface plasmon polaritons (SPPs)~\cite{RN33}, and realized the artificial domain wall that supports highly confined one-dimensional plasmonic modes.
	
	Motivated by these studies, in this paper, we propose a transfer matrix method (TMM) to study the GPs under periodic conductivity by applying an appropriate gate voltage for a monolayer graphene. Following Ref.~\cite{PhysRevB.90.041407}, we construct the transfer matrix of a jumping interface from the transmission and reflection coefficients. Graphene, whose Fermi levels are periodically arranged along the plane caused by periodic electric potential, can be regarded as a conductivity plate that periodically jumps and its rich optical properties can be easily studied by TMM. This will serve as an efficient tool for numerical simulation. The effectiveness of the TMM, introduced in our model, is also verified with the plane wave expansion method. Furthermore, we achieve defect induced transparency due to the presence of a defect in the GPC and also use the GPC to achieve plasmonic Tamm state.
	
	The paper is organized as follows: In Sec. II, we give our model and analytical description. The results for the transmission characteristics of the GPC, both in the presence as well as in the absence of the defect, and for achieving plasmonic tamm state are discussed in Sec. III. In the last section (Sec. IV) we conclude the study.
	
	\section{\label{sec:level2}Model and simulation}
	We start our model by properly addressing the optical properties of graphene, being the most important component of our scheme. In the long wavelength and high doping limit, i.e., $\hbar\omega \ll E_{F}$, the in-plane conductivity of graphene can be described by a simple semiclassical Drude's model $\sigma(\omega)=i e^{2}E_{F}/[\pi\hbar^{2}(\omega+i/\tau)]$ under the random-phase approximation~\cite{hwang2007dielectric,jablan2009plasmonics}. Here $E_{F}$ is the Fermi energy and $\tau$ is the momentum relaxation time due to impurity or phonon-mediated scattering.
	
	Similar to normal metals having free electrons' collective oscillation, the electron carriers of the doped graphene can also respond to the electromagnetic field resonantly leading to graphene plasmons (GPs). For a plane dielectric-monolayer graphene-dielectric model, the plasmon dispersion relation has the following form~\cite{zeng2014nanometer,hwang2007dielectric,gullans2013single}
	\begin{equation}\label{k}
		k_{p}\approx\frac{(\epsilon_1+\epsilon_2)}{4\alpha}\frac{\omega\hbar}{E_F}k_0\left(1+\frac{i}{\tau\omega}\right),
	\end{equation}
	where $\alpha=e^{2}/(4\pi\hbar\epsilon_0c)\approx1/137$ is the fine-structure constant and $k_0$ is the vacuum wave number. Here, $\epsilon_1$=1 corresponding to air and $\epsilon_2$=2.25 corresponding to silicon dioxide spacers are the relative dielectric permittivities above and below the graphene. Note that the magnitude of GPs' wave vector will change with the change of the Fermi energy $E_F$, which means that we can get the wave vector we need by adjusting the gate voltage.
	\begin{figure}[htbp]
		\centering
		\includegraphics[width=\linewidth]{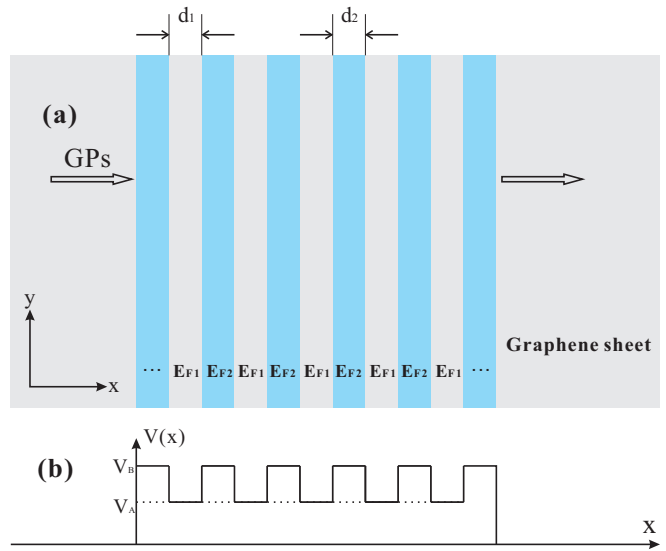}\\
		\caption{(a) The infinite graphene sheet is on $z=0$ plane, the bottom ($z<0$) is the silicon dioxide spacers, and the top ($z>0$) is air. The grey and blue areas represent the Fermi energy levels corresponding to $E_{F1}$ and $E_{F2}$. We assume that the width of the graphene region corresponding to each bias gate voltage is the same, that is,  $d_1=d_2=d$. GPC is formed when GPs pass through periodic Fermi energy regions. (b) The profiles of the periodic gate voltage applied on the monolayer graphene. $V_A$ and $V_B$ correspond respectively to $E_{F1}$ and $E_{F2}$.}
		\label{FIG1}
	\end{figure}
	When considering the periodic gate voltage distribution of the A-B-A-B type, as shown in Fig. \ref{FIG1}(b), this distribution will cause the Fermi energy of graphene to be periodically distributed. It leads to a periodic conductivity distribution because the Fermi energy is proportional to the conductivity. Unlike the traditional photonic crystals with periodic $\epsilon(x)$, GPCs form the lattice structure through the distribution of position-dependent conductivity $\sigma(x)$. Further, considering the surface plasmons (p-polarized wave) incident along the x-axis into the GPC, under the quasi-static approximation, the equation for the incident wave induced surface current behavior can be written as~\cite{Rejaei_2015}
	\begin{equation}\label{self-consistent}
		\dfrac{j_x(x)}{\sigma(x)}=E^{ex}_x(x)+\dfrac{1}{2\pi i \omega\epsilon_0\epsilon_e}\int^{\infty}_{-\infty}\dfrac{\partial_{x'} j_x(x')}{x'-x}dx',
	\end{equation}
	where $\epsilon_e$=($\epsilon_1$+$\epsilon_2$)/2. The integral is of the Cauchy principal-value type. Considering the GPCs in one period, that is, the A-B region, the reflection and transmission coefficients of surface current when passing through the jumping interface $i|j$ are obtained by the Wiener-Hopf method as~\cite{Rejaei_2015}
	\begin{equation}\label{eq3}
		r_{ij}=e^{i\theta_{ij}}\dfrac{k_i-k_j}{k_i+k_j},\quad
		t_{ij}=t_{ji}=\dfrac{2\sqrt{k_ik_j}}{k_i+k_j},
	\end{equation}
	where $k_{i,j}$ is the local value of the wave number of GPs at two different Fermi levels. The anomalous reflection phase $\theta_{ij}$ in Eq. (\ref{eq3}) is given by
	\begin{equation}
		\theta_{ij}=-\theta_{ji}=\dfrac{\pi}{4}-\dfrac{2}{\pi}\int^{\infty}_{0}\dfrac{\arctan(k_ix/k_j)}{1+x^2}dx.
	\end{equation}
	The detailed derivation of the above equations is given in Appendix \ref{Appendix A}. 
	It is worth mentioning that the analytical results obtained by this method are only applicable within a certain range. During the scattering process across the jumping interface, many different modes are excited. These modes travel a certain distance and then couple to GPs~\cite{PhysRevB.90.041407}. With this in mind, the value chosen for the width $d$ in our model is prudent and long enough.
	
\begin{figure}[htbp]
		\centering
		\includegraphics[width=\linewidth]{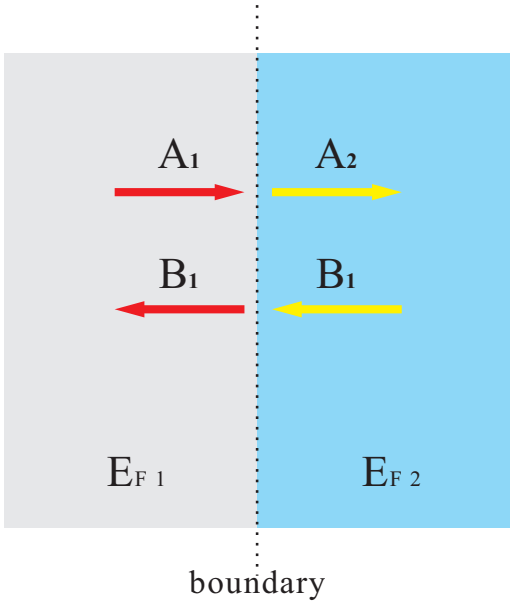}\\
		\caption{For both sides of a single boundary, its field is composed of the front wave and back wave of GPs. The red arrows represent the GPs under Fermi energy $E_{F1}$ and the yellow arrows represent the GPs under Fermi energy $E_{F2}$. }
		\label{FIG2}
	\end{figure}

	Now we have the reflectance and transmittance of a single jumping interface, which is the basis for us to introduce TMM to study the GPCs. The total field adjacent to both sides of the jumping interface can be written in the following form (where we have assumed that all modes have been quickly coupled to GPs)
	\begin{align}\label{eq5}
		\begin{split}
			E_L(x=0^-)=A_1e^{ik_1x}+B_1e^{-ik_1x}, \\
			E_R(x=0^+)=A_2e^{ik_2x}+B_2e^{-ik_2x}. 	
		\end{split}
	\end{align}
	Here, $A_{1,2}$ and $B_{1,2}$ represent the amplitudes of the GPs propagating along the positive and negative x-axes on both sides, as shown in Fig. \ref{FIG2}.
	
	The matrix $M_{12}$ is introduced, connecting $A_1$ and $B_1$ to $A_2$ and $B_2$
	\begin{equation}\label{eq6}
		\begin{pmatrix}
			A_1\\B_1
		\end{pmatrix}
		=
		M_{12}
		\begin{pmatrix}
			A_2\\B_2
		\end{pmatrix}
		=
		\begin{pmatrix}
			M^{(1)}_{12} & M^{(2)}_{12}\\
			M^{(3)}_{12} & M^{(4)}_{12}
		\end{pmatrix}
		\begin{pmatrix}
			A_2\\B_2
		\end{pmatrix}
		.
	\end{equation}
	In order to solve the matrix $M_{12}$, here we introduce two independent surface currents, $J_1$ and $J_2$. Assuming that $J_1$ and $J_2$ are incident independent of each other in opposite directions from both sides of the middle boundary of a single period. The two currents scatter at the interface as described by Eq. (\ref{eq3}). Therefore,  Eq. (\ref{eq5}) can also be written as
	\begin{equation}\label{eq7}
		\begin{aligned}
			E_L(x=0^-)=&\dfrac{J_1}{\sigma_1}e^{ik_1x}+\left(\dfrac{J_1}{\sigma_1}r_{12}+\dfrac{J_2}{\sigma_1}t_{21}\right)e^{-ik_1x}, \\
			E_R(x=0^+)=&\left(\dfrac{J_1}{\sigma_2}t_{12}+\dfrac{J_2}{\sigma_2}r_{21}\right)e^{ik_2x}+\dfrac{J_2}{\sigma_2}e^{-ik_2x},	
		\end{aligned}
	\end{equation}
	where $\sigma_1$ and $\sigma_2$ are the local conductivitivities under Fermi energy $E_{F1}$ and $E_{F2}$. $J_1$ is the current propagating to the right on the left side of the interface and $J_2$ is the current propagating to the left on the right side of the interface. Comparing Eqs. (\ref{eq5}) and (\ref{eq7}), we get
	\begin{equation}
		\begin{aligned}\label{eq8}
			A_1&=\dfrac{J_1}{\sigma_1},                 &
			A_2&=\dfrac{J_1}{\sigma_2}t_{12}+\dfrac{J_2}{\sigma_2}r_{21},\\
			B_1&=\dfrac{J_1}{\sigma_1}r_{12}+\dfrac{J_2}{\sigma_1}t_{21}, &
			B_2&=\dfrac{J_2}{\sigma_2}.
		\end{aligned}
	\end{equation}
	\\
	While comparing Eqs. (\ref{eq6}) and (\ref{eq8}), we obtain  
	\begin{equation}\label{eq9}
		\begin{aligned}
			\dfrac{J_1}{\sigma_1}&=M^{(1)}_{12}\left(\dfrac{J_1}{\sigma_2}t+\dfrac{J_2}{\sigma_2}r_{21}\right)+M^{(2)}_{12}\dfrac{J_2}{\sigma_2},\\
			\dfrac{J_1}{\sigma_1}r_{12}+&\dfrac{J_2}{\sigma_1}t=M^{(3)}_{12}\left(\dfrac{J_1}{\sigma_2}t+\dfrac{J_2}{\sigma_2}r_{21}\right)+M^{(4)}_{12}\dfrac{J_2}{\sigma_2},
		\end{aligned}
	\end{equation}
	where we use $t_{12}=t_{21}=t$. Each element of the matrix $M_{12}$ can be solved by the linear independence of $J_1$ and $J_2$ in Eq. (\ref{eq9}), so $M_{12}$ yields
	\begin{equation}\label{eq10}
		M_{12}=\dfrac{\sigma_2}{\sigma_1}
		\begin{pmatrix}
			1/t & -r_{21}/t\\
			r_{12}/t & 1/t
		\end{pmatrix}
		,
	\end{equation}
	where the relationship $t_{12}t_{21}-r_{12}r_{21}$=1 is used. In the same way, we can obtain the matrix $M_{21}$ as
	\begin{equation}\label{eq11}
		M_{21}=\dfrac{\sigma_1}{\sigma_2}
		\begin{pmatrix}
			1/t & r_{21}/t\\
			-r_{12}/t & 1/t
		\end{pmatrix}
		.
	\end{equation}
	Note that the relationship $M_{12}$=$M_{21}^{-1}$ can be easily verified. Except for the behavior of the GPs at the jumping boundary, the front and the back waves within the adjacent boundary can also be related by the matrix
	\begin{equation}\label{eq12}
		P_{1,2}=
		\begin{pmatrix}
			e^{-ik_{1,2}d} & 0\\
			0 & e^{ik_{1,2}d}
		\end{pmatrix}
		,
	\end{equation}
	where the subscripts of $P_{1,2}$ correspond to the Fermi level $E_{F1}$ and $E_{F2}$, respectively.
	
\begin{figure}[htb]
		\centering
		\includegraphics[width=\linewidth]{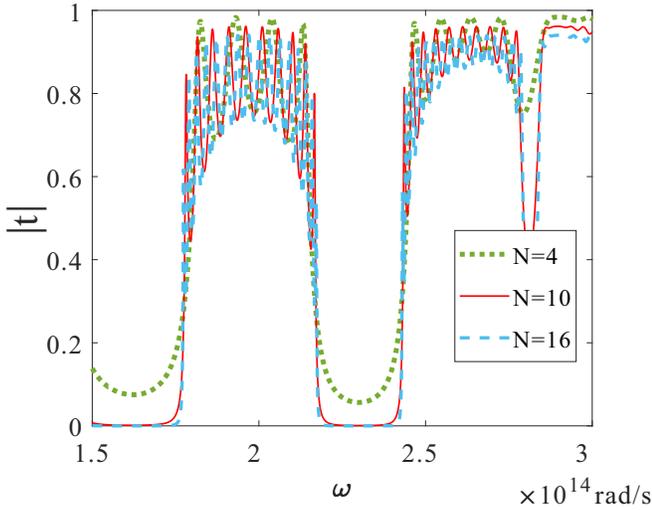}\\
		\caption{Transmission coefficient diagram of the GPC where $d$=100nm, $\tau$=10$^{-11}$rad/s, $E_{F1}$=0.3eV and $E_{F2}$=0.65eV. The green dotted line corresponds to $N$=4, the red line corresponds to $N$=10, and the blue dashed line corresponds to $N$=16.}
		\label{FIG3}
	\end{figure}

	For one period, from the left side of $E_{F1}$ to the left side of the next $E_{F1}$, the transfer matrix element $M_0$ connecting its field magnitude can be obtained as
	\begin{widetext}
		\begin{equation}\label{eq13}
			\begin{aligned}
				M_{0}=M_{12}\cdot P_2\cdot M_{21}\cdot P_1
				=t^{-2}
				\begin{pmatrix}
					r_{12}r_{21}e^{i(k_2-k_1)d}+e^{-i(k_1+k_2)d} &
					r_{21}(e^{i(k_1-k_2)d}-e^{i(k_1+k_2)d})        \\
					r_{12}(e^{-i(k_1+k_2)d}-e^{i(k_2-k_1)d}) &
					r_{21}r_{12}e^{i(k_1-k_2)d}+e^{i(k_1+k_2)d}
				\end{pmatrix}
				.
			\end{aligned}
		\end{equation}
	\end{widetext}
	With this in hand, the transfer matrix of a GPC with N periods can be written as
	\begin{equation}\label{eq14}
		\begin{aligned}
			M=M_{0}^{N}\cdot M_{12} \cdot P_2 \cdot M_{21}=
			\begin{pmatrix}
				M^{(11)} &
				M^{(12)}        \\
				M^{(21)} &
				M^{(22)}
			\end{pmatrix}
			,
		\end{aligned}
	\end{equation}
	and the transmission and reflection coefficients  of the GPC can be easily obtained as
	\begin{equation}\label{eq15}
		\begin{aligned}
			t=\dfrac{1}{M^{(11)}},\\
			r= \dfrac{M^{(21)}}{M^{(11)}}.
		\end{aligned}
	\end{equation}
	The number of periods of the GPC can be theoretically infinite but usually it only needs to be set to a finite value, which is also sufficient to reflect the properties of a GPC. As shown in Fig. \ref{FIG3}, the transmission coefficient spectrum of the GPC under different number of periods is calculated by TMM. 
	We see that when the number of periods is 6, compared to the higher number of $N$, the complete forbidden band ($|t|$=0) does not appear. When the number of periods increases to 10, the GPC already has an obvious forbidden band and there is almost no change compared with $N$=16. This is in accordance with the properties of traditional photonic crystals but the smaller size of the GPC is more advantageous. In the following, the number of periods of the GPC simulated by TMM is all set to be $N$=10.
	
	It is worth mentioning that when the frequency is higher than 0.2eV, the phonon mode generated by graphene will couple with GPs, thereby affecting the GPs. Therefore, in our simulation, the photon frequency range is kept below 0.2eV. In actual situation, when GPs pass through the jumping interface, several other modes will be excited but they are all short-lived. This will cause the actual phase accumulation in the gate to be different from $k_{p} d$. In our simulation, the value of $d$ is taken to be 100nm and we can approximately treat the phase accumulated in the gate as being accumulated by local GPs~\cite{PhysRevB.90.041407}.
	
	\begin{figure}[htbp]
		\centering
		\includegraphics[width=\linewidth]{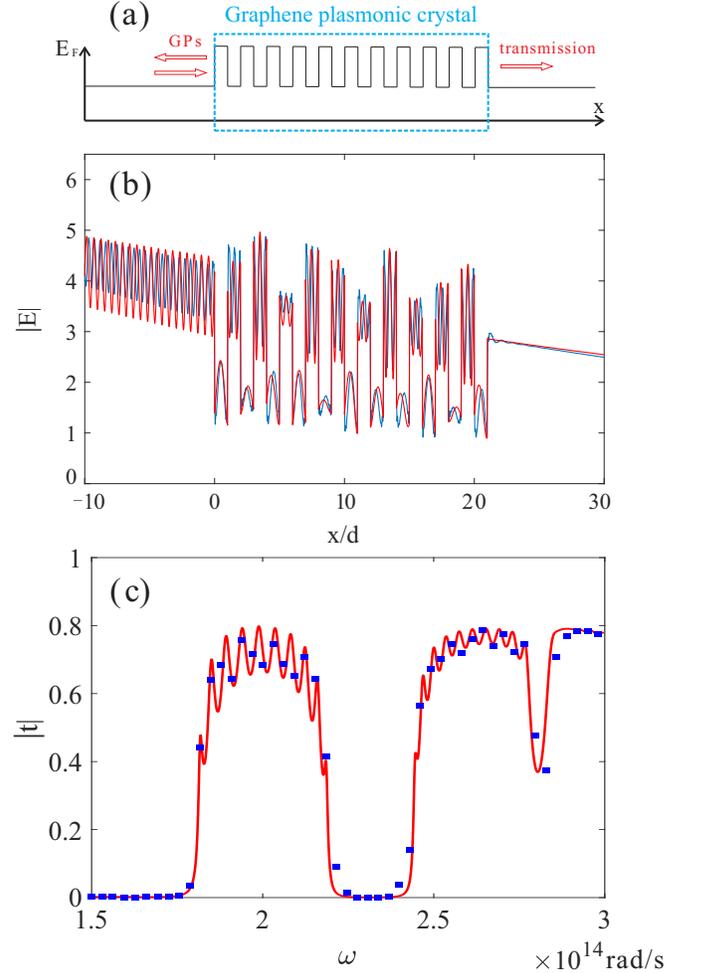}\\
		\caption{(a) The Fermi energy distribution of graphene, where higher energy corresponds to $E_{F2}$ and lower energy corresponds to $E_{F1}$. The blue dashed area correspond to GPC. GPs enter from the left side of the GPC and exit from the right end. (b) The field distribution of an incident GP through the GPC such that $\omega/\omega_0$=1.4, $E_{F1}$=0.3eV, $E_{F2}$=0.65eV, $d$=100nm and $\tau$=1/6$\times10^{-11}$rad/s. (c) The transmission coefficient of the GPC with same set of parameters. Both in (b) and (c), the red lines show the results obtained via TMM and the blue lines correspond to the results obtained via plane wave expansion method.}
		\label{FIG4}
	\end{figure}
	In order to verify the effectiveness of TMM, we use the plane wave expansion method to numerically simulate the plasmonic behavior of the GPC. The specific introduction of this method is given in Appendix \ref{appendixB}. In actual simulations, the application range of plane wave expansion method will be wider because of the accuracy of its intrinsic formula, but in this structure TMM can save more time and memory than plane wave expansion method or boundary element method. The results obtained for the plasmonic behavior of the GPC via TMM and plane wave expansion method are compared in Fig.~\ref{FIG4}.
	In Fig.~\ref{FIG4}(b), we use plane wave expansion method to simulate the field distribution of an incident field when it passes through the GPC. In regions where $x<0$, an obvious standing wave can be observed which is caused by the interference effect between the incident and the reflected GPs. 
	In GPC region, standing waves can also be seen and the number of standing waves corresponds to the wavelength of the local GPs ($\lambda_{1}$=93.86nm, $\lambda_{2}$=203.37nm). On the right side of the GPC, that is, in regions $x>21d$, we can see that under plane wave expansion method simulation, the modes outgoing the GPC are quickly coupled to GPs which also confirm the effectiveness of the parameters selected.
	In Fig.~\ref{FIG4}(c), the transmission coefficient of the GPC is independently simulated by TMM (red line) and plane wave expansion method (blue dotted line). It can be seen that the results of these two simulations are consistent.
	
	\begin{figure}[htbp]
		\centering
		\includegraphics[width=\linewidth]{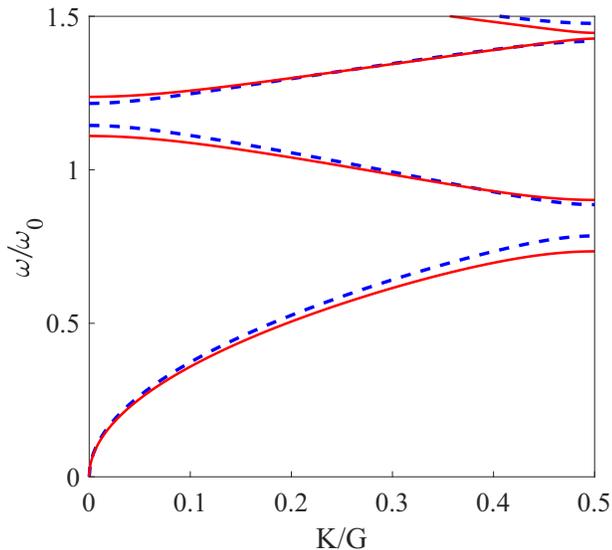}\\
		\caption{The band structure of the GPC calculated by TMM (red line) and plane wave expansion method (blue dashed line). Here $E_{F1}$=0.3eV, $E_{F2}$=0.65eV, $d$=100nm and $G=\pi/d$ is the reciprocal lattice vector.}
		\label{FIG5}
	\end{figure}
	Besides the transmission characteristics and field distribution  , the dispersion relation of the GPC can also be simply obtained from the TMM. According to Bloch theorem, the field at the periodic interface satisfies
	\begin{equation}\label{eqBand1}
		\begin{aligned}
			\begin{pmatrix}
				A
				\\
				B
			\end{pmatrix}=
			\begin{pmatrix}
				e^{-iK\varLambda} &
				0\\
				0  &
				e^{iK\varLambda} 
			\end{pmatrix}
			\cdot
			\begin{pmatrix}
				A'      \\
				B'
			\end{pmatrix}
			,
		\end{aligned}
	\end{equation}
	where $\varLambda$ is the period of the GPC and $K$ is effective wave vector.
	Then, from Eq. (\ref{eq14}) we get
	\begin{equation}\label{eqBand2}
		\begin{aligned}
			\begin{pmatrix}
				A
				\\
				B
			\end{pmatrix}=
			M_0 \cdot
			\begin{pmatrix}
				A'      \\
				B'
			\end{pmatrix}=
			\begin{pmatrix}
				M_0^{(11)} & M_0^{(12)}     \\
				M_0^{(21)} & M_0^{(22)}
			\end{pmatrix} \cdot
			\begin{pmatrix}
				A'      \\
				B'
			\end{pmatrix}
			,
		\end{aligned}
	\end{equation}
	Comparing Eqs. (\ref{eqBand1}) and (\ref{eqBand2}), the dispersion relation of the GPC takes the form
	\begin{equation}\label{eq18}
		\cos(K\varLambda)=\dfrac{1}{2}\left(M_0^{(11)}+M_0^{(22)}\right).
	\end{equation}
	In Fig. \ref{FIG5}, the band gap structures of the GPC calculated by Eq.~(\ref{eq18}) and plane wave expansion method are plotted.
	Here $\omega_0$ is given by
	\begin{equation}
		\omega_0=\sqrt{\dfrac{e^2E_{F0}}{\epsilon_0\hbar^2L}},
	\end{equation} 
	with $L$=100nm and $E_{F0}$=0.1eV.
	We can see that the band gap structures calculated by these two methods are almost the same. 
	
	\begin{figure}[htb]
		\centering
		\includegraphics[width=\linewidth]{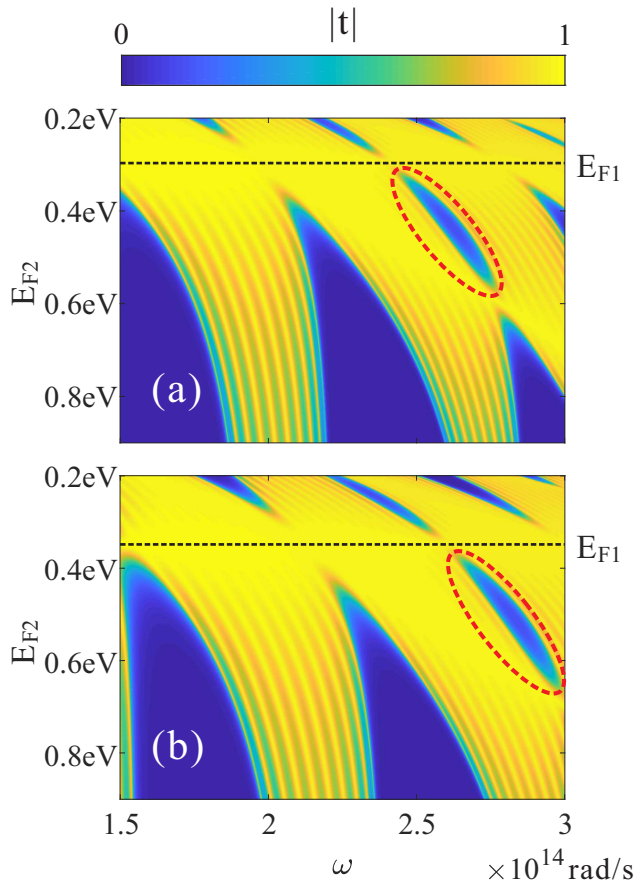}\\
		\caption{Transmission coefficient diagram of the GPC for varying $E_{F2}$ and (a) $E_{F1}$=0.3eV while (b) $E_{F1}$=0.35eV. The black dotted line is the cutoff line with $E_{F2}$=$E_{F1}$. The red dashed line marks the contrast area. Other parameters used are $d$=100nm and $\tau$=5$\times10^{-12}$rad/s.}
		\label{FIG6}
	\end{figure}
	GPC has better tunability as compared to traditional photonic crystals. When the platform is encapsulated, the conductivity distribution of graphene can be adjusted by adjusting the gate voltage, thereby varying the photonic properties of the GPC. As shown in Fig. \ref{FIG6}, for a fixed $E_{F1}$ and varying $E_{F2}$, different transmission characteristics are noted for the GPC under TMM simulations. 
	When $E_{F2}$ becomes equal to $E_{F1}$ (the black dashed line in the Fig. \ref{FIG6}), the GPC becomes a graphene sheet with uniform conductivity and is almost transparent, which is also in agreement with the properties of GPs (long lifetime). Both Figs. \ref{FIG6}(a) and \ref{FIG6}(b) show that by adjusting the Fermi level $E_{F2}$, some narrow or wide band gaps can be obtained. For example, in the red dashed area, the forbidden band is very narrow which can be used as an absorber in a specific frequency domain. In some areas, the band gap is wider and can be used as a filter in a specific frequency domain.
	
	\begin{figure}[htbp]
		\centering
		\includegraphics[width=\linewidth]{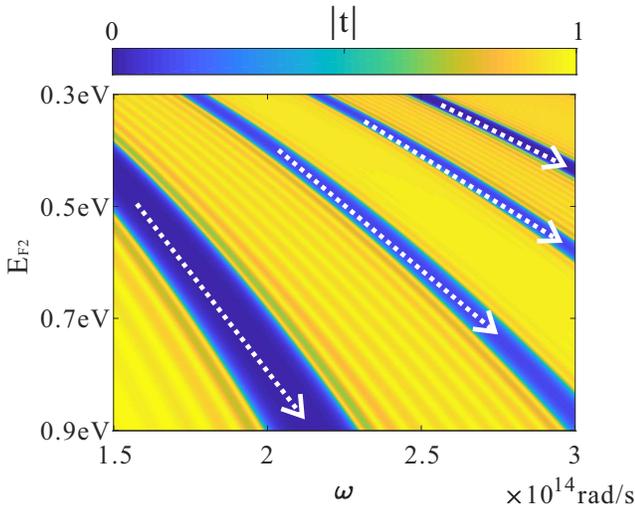}\\
		\caption{Transmission coefficient diagram of the GPC where $E_{F2}/E_{F1}$ remains fixed at 1.5. The white dotted lines mark the evolution direction of each forbidden band when $E_{F2}$ increases. Other parameters used are $d$=100nm and $\tau$=5$\times10^{-12}$rad/s.}
		\label{FIG7}
	\end{figure}
	Comparing Figs. \ref{FIG6}(a) and \ref{FIG6}(b), when $E_{F1}$ is increased from 0.3eV to 0.35eV, it can be found that the photonic transmission characteristics shift to the lower right direction (it can be observed by comparing the red dashed area). This means that the transmission characteristics realized in a fixed frequency domain can also be observed in different frequency domains by adjusting the Fermi level of graphene. This property is also reflected in Fig. \ref{FIG7}, where the transmission coefficient diagram of the GPC is calculated by TMM at a fixed ratio $E_{F2}/E_{F1}=1.5$. When the ratio of the Fermi energies $E_{F2}$ to $E_{F1}$ remains constant, the position of the band gaps will move to the right as $E_{F2}$ increases, and the width of the band gap remains almost unchanged.
	
	As we know that the lifetime of GPs can be described by the propagation length, which is defined to be 
	\begin{equation}\label{eq25}
		L=\dfrac{1}{2Im\{k_{GP}\}}.
	\end{equation}
	Comparing this with Eq. (\ref{k}) it is not difficult to find 
	\begin{equation}\label{eq26}
		L\propto \dfrac{E_F}{\omega}\equiv\varrho.
	\end{equation}
	This means that, while keeping $\varrho$ constant, the lifetime of GPs will not be affected. Therefore, by adjusting $E_F$, the GPC can maintain a good lifetime for GPs even when the $\omega$ is large. This reflects a useful property of the GPC. In Fig. \ref{FIG7}, after a simple estimation, it is not difficult to find that along the directions of the white dotted lines (where the photonic characteristic remains unchanged), the value of $\varrho$ increases as the $\omega$ increases. Then we can conclude that when $\omega$ and $E_{F1}$ or $E_{F2}$ are proportional to each other, the GPC can achieve similar photonic transmission characteristics under different frequency domain and the lifetime of the GPs will increase with the increase of $\omega$. 
	
	\section{\label{sec:level3}Defect mode and plasmonic tamm state }
\begin{figure}[htb]
		\centering
		\includegraphics[width=\linewidth]{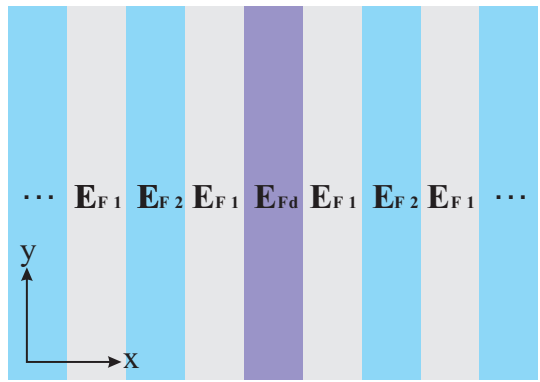}\\
		\caption{The top view of a GPC structure with defect introduced. The Fermi energy of the defect is $E_{Fd}$ corresponding to the purple area, replacing the original Fermi energy $E_{F2}$.}
		\label{FIG8}
	\end{figure}

	In the following, we propose a defect mode based on GPC, which can be modified by changing the Fermi energy of the defect area, as shown in Fig. \ref{FIG8}. 
	The wave vector of the local GPs in the defect is $k_d$, which is obtained by Eq. (\ref{k}). Then, analogous to Eq. (\ref{eq14}) and the previous derivation, we can write the transfer matrix of the GPC that introduces the defect as
	\begin{equation}\label{M_d}
		\begin{aligned}
			M_d=M_{0}^{N/2}\cdot M_{1d} \cdot P_d \cdot M_{d1} \cdot P_1 \cdot M_{0}^{N/2} \cdot M_{12} \cdot P_2 \cdot M_{21},
		\end{aligned}
	\end{equation}
	where we assume that $N$ is even. $M_{1d}$, $M_{d1}$ and $P_d$ are given by
	\begin{align}\label{M1d_Md1}
		M_{1d}=M_{d1}^{-1}=&\dfrac{\sigma_d}{\sigma_1}
		\begin{pmatrix}
			1/t_{1d} & -r_{d1}/t_{1d}\\
			r_{1d}/t_{1d} & 1/t_{1d}
		\end{pmatrix}
		,\\
		P_{d}=&
		\begin{pmatrix}
			e^{-ik_{d}d} & 0\\
			0 & e^{ik_{d}d}
		\end{pmatrix}
		.
	\end{align}
	The transmission coefficient of the GPC can be obtained as
	\begin{equation}\label{TR_defect}
		\begin{aligned}
			t_d=\dfrac{1}{M_d^{(11)}}.
		\end{aligned}
	\end{equation}
	\begin{figure*}[htbp]
		\centering
		\includegraphics[width=6.5in]{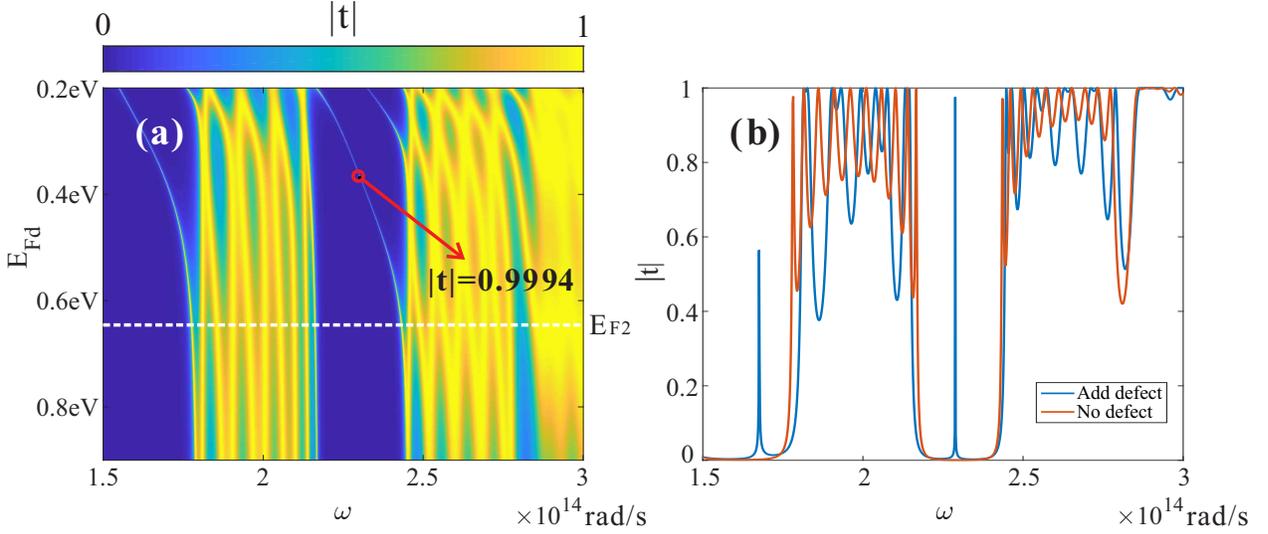}\\
		\caption{(a) Transmission coefficient spectrum of the GPC with defect introduced in it. Here, $E_{F1}$=0.3eV $E_{F2}$=0.65eV and $d$=100nm.  The Fermi energy of the defect $E_{Fd}$ varies from 0.2eV to 0.9eV. The white dashed line corresponds to $E_{Fd}$=$E_{F2}$=0.65eV and the red circles with pointed arrows represent the transmission coefficient of a point on the transparent band gap caused by the defect. (b) Transmission coefficient spectrum of the GPC. The red line corresponds to the GPC without defect, that is, $E_{Fd}$=$E_{F2}$=0.65eV, and the blue line corresponds to the GPC with defect, that is, $E_{Fd}$=0.34eV$\neq E_{F2}$. The other parameters of the red and blue lines are the same, $d$=100nm, $\tau^{-1}$=0.}
		\label{FIG9}
	\end{figure*}\\
	As shown in Fig. \ref{FIG9}(a), for a fixed $E_{F1}$, $E_{F2}$ and varying $E_{Fd}$, transmission coefficient spectrum of GPC is calculated by Eq. (\ref{TR_defect}). In order to clearly observe the transparency of GPC caused by the defect, we assume that there is no dissipation in the model, that is, $\tau\rightarrow\infty$.  When the Fermi energy of the defect $E_{Fd}$ is equal to $E_{F2}$, the model is equivalent to normal GPC, corresponding to the white dashed line in Fig. \ref{FIG9}(a). The comparison of the transmission coefficient spectrum of GPC with and without defect is plotted in Fig \ref{FIG9}(b). The red line corresponds to the normal GPC, and the band gap prohibits photons from passing through the GPC, that is $|t|$=0. As shown by the blue line, when the defect is introduced in the GPC, transparency appears in the band gap, that is, $|t|\approx$1. This shows that we can adjust the Fermi energy of the defect to make the original forbidden band transparent, and the width of the transparent band is narrow. 
	The introduction of the defect mode provides a higher degree of freedom for the GPC, and the transparency guided by the defect provides a tool for the realization of advanced physics concepts. 
	
\begin{figure}[htb]
		\centering
		\includegraphics[width=\linewidth]{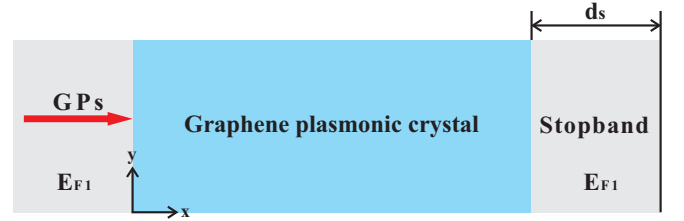}\\
		\caption{The top view of the GPC-Stopband structure which is infinite in the y direction. Here, a stopband is added to the right side of the GPC. The Fermi energy of the stopband is selected as $E_{F1}$ and its width is $d_s$. The GPs, that correspond to Fermi level $E_{F1}$, enter from the left side of the GPC. }
		\label{FIG10}
	\end{figure}

	In addition, a GPC can be used as a platform for plasmonic Tamm states and the results can be easily simulated with TMM. In order to build a plasmonic Tamm state, a stopband is added to the end of the GPC as shown in Fig.~\ref{FIG10}. The GPs in this model will reflect strongly at the end of the stopband because of the terminal which prevents the spread of energy into radiating mode and leading to highly confined surface plasmons.
	At the right end of the stopband, which is also the edge of graphene, its transmittance for GPs (transmitted into the air) is almost zero, and this can be confirmed by Eq. (\ref{eq3}) (when $k_j\rightarrow\infty$, $r_{ij}\approx e^{-i3\pi/4}$, $t_{ij}$=0). Therefore, at the right end of the stopband, the amplitude of the front and back propagating GPs can be assumed as
	\begin{equation}\label{eq20}
		\begin{pmatrix}
			A_s        \\
			B_s
		\end{pmatrix}
		=
		\begin{pmatrix}
			1        \\
			r_s
		\end{pmatrix},
	\end{equation}
	where $r_s$=$e^{-i3\pi/4}$. Then the amplitude of the GPs at the entrance (left end) of the GPC is given by
	\begin{equation}\label{eq21}
		\begin{pmatrix}
			A_{0}        \\
			B_{0}
		\end{pmatrix}
		=M_{GPC}P_s
		\begin{pmatrix}
			1        \\
			r_s
		\end{pmatrix},
	\end{equation}
	where $M_{GPC}$ is the transfer matrix of the GPC, which can be obtained from Eq. (\ref{eq14})
	\begin{equation}\label{eq22}
		M_{GPC}=M_0^N\cdot M_{12}\cdot P_2 \cdot M_{21},
	\end{equation} 
	and $P_s$ is 
	\begin{equation}\label{eq23}
		P_s
		=
		\begin{pmatrix}
			e^{-ik_1d_s} &  0      \\
			0            &  e^{ik_1d_s}
		\end{pmatrix}.
	\end{equation}
	The total reflection coefficient of the GPC-Stopband structure is 
	\begin{equation}\label{eq24}
		r_{tot}=\dfrac{B_{0}}{A_{0}}.
	\end{equation}
	Note that since TMM itself considers the accumulation of phase, when using this method to obtain the total reflection coefficient of the GPC-Stopband structure, the phase matching condition will be automatically satisfied~\cite{PhysRevApplied.12.024057}.
	
	Considering the generation of the Tamm state, in addition to studying the reflection properties of the structure, the electromagnetic field distribution is also an important basis for determining whether there is a Tamm state generated. The transfer matrix can be regarded as a connection of the fields at each position on the graphene sheet. When the magnitude of the front and back waves somewhere on the graphene surface are determined, the field distribution of the entire plane can be easily obtained by TMM.
	
	\begin{figure}[tp]
		\centering
		\includegraphics[width=\linewidth]{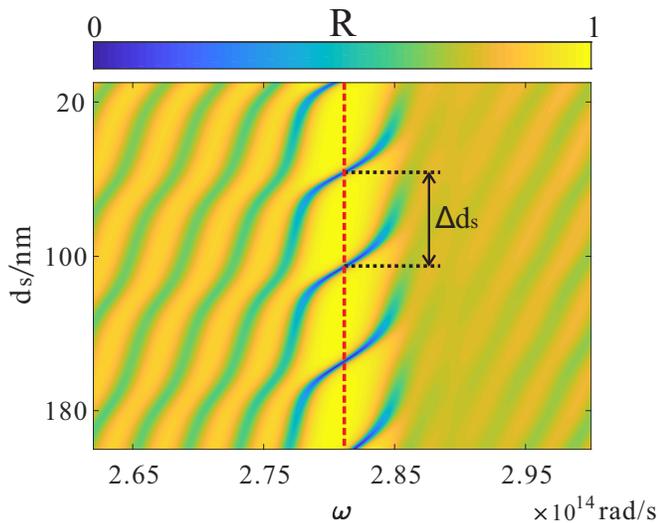}\\
		\caption{The reflectance spectrum of the GPC-Stopband structure where $d_s$ varies from 10nm to 200nm. Other parameters used are $E_{F1}$=0.3eV, $E_{F2}$=0.65eV, $d$=100nm and $\tau$=5$\times10^{-12}$rad/s. The red dashed line corresponds to $\omega_m$=2.81$\times10^{14}$rad/s.}
		\label{FIG11}
	\end{figure}
	For the GPC-Stopband structure in Fig. \ref{FIG10}, the stopband at the tail is equivalent to introducing additional phase and returning energy to the GPC. The reflection of energy is determined by $|r_s|^2$ and the phase compensation is determined by its width $d_s$ and local GPs. The reflectance spectrum of the structure in the range from $d_s$=10nm to $d_s$=200nm calculated by Eq. (\ref{eq24}) is plotted in Fig. \ref{FIG11}. The red dashed line corresponds to $\omega_m$=2.81$\times10^{14}$rad/s and the minimum and maximum value of $R$ are on this dashed line. It can be seen that $R$ has a higher contrast ratio at $\omega_m$ than other frequencies. It is worth mentioning that, because $d_s$ determines the compensation phase, the figure shows periodicity. It can be easily verified that the period $\Delta d_s$ at $\omega_m$ has the following relationship
	\begin{equation}\label{eq27}
		\Delta d_sk_1=\pi,
	\end{equation}
	where $k_1$ is the wave vector of local GPs at $\omega_m$ corresponding to Fermi energy $E_{F1}$ in the stopband.
	
	\begin{figure}[htp]
		\centering
		\includegraphics[width=\linewidth]{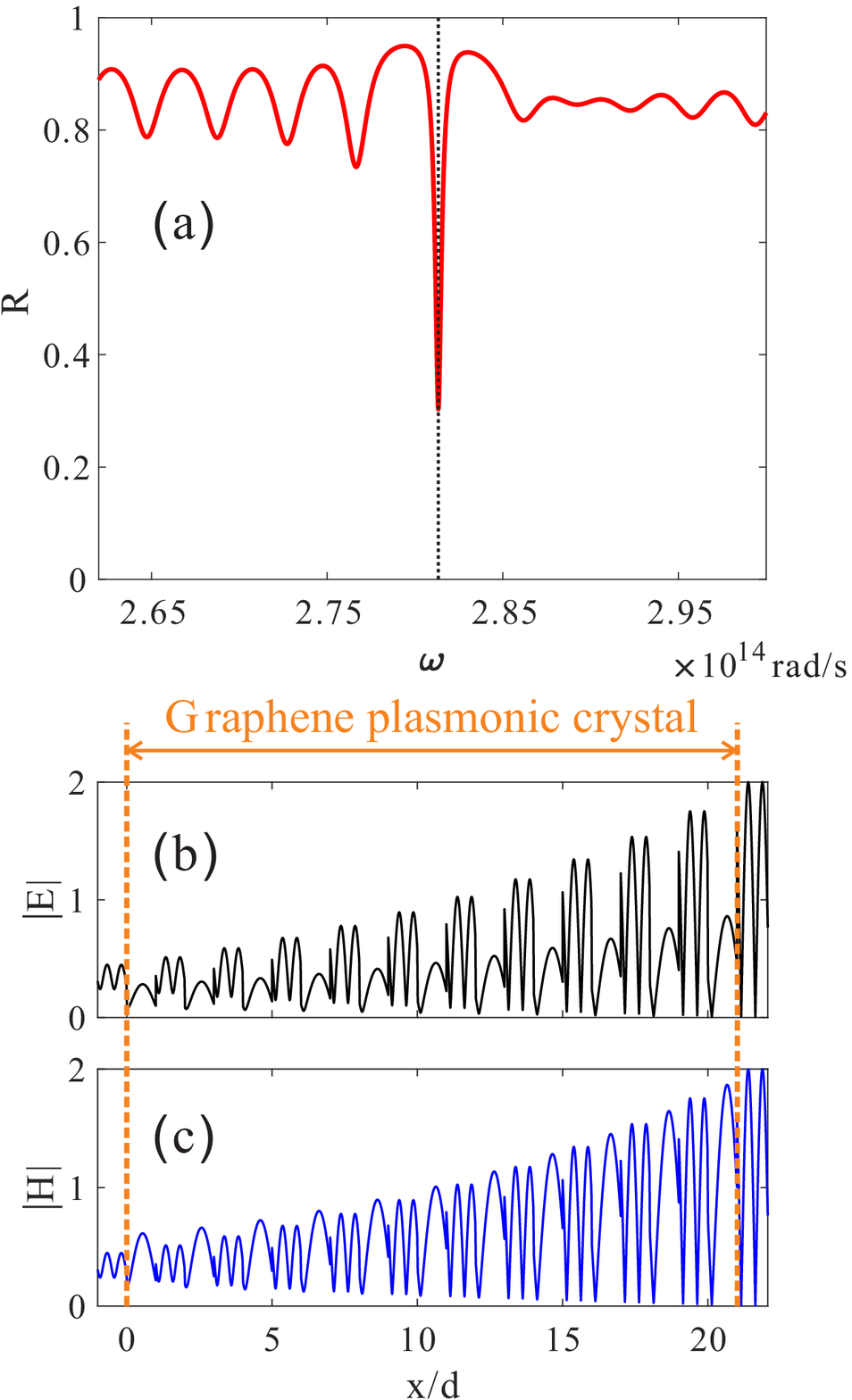}\\
		\caption{(a) The reflectance spectrum of the GPC-Stopband structure. The dotted line corresponds to $\omega$=2.81$\times10^{14}$rad/s, $E_{F1}$=0.3eV, $E_{F2}$=0.65eV,  $d$=100nm, $\tau$=5$\times10^{-12}$rad/s and $d_{s}$=106nm. (b),(c)  The field distributions of the normalized electric and magnetic fields at $\omega$=2.81$\times10^{14}$rad/s. The area $x<0$ correspond to Fermi energy $E_{F1}$ graphene plane, the yellow dashed area, that is from $x=0$ to $x=21d$ , is the GPC, and the area from $x=21d$ to $x=22.06d$ represents the stopband. }
		\label{FIG12}
	\end{figure}
	The sharp dip in the reflectance spectrum is a prerequisite for plasmonic Tamm state. Around $d_s$=100nm (other values can also be selected), the reflectance has a minimum at $d_s$=106nm. The reflectance spectrum of the GPC-Stopband structure calculated by TMM under this $d_s$ is plotted In Fig. \ref{FIG12}(a). Here, the minimum value is $30.29\%$ with a full width at half maximum of only 0.08THz. 
	Figures \ref{FIG12}(b) and \ref{FIG12}(c) show the normalized electric and magnetic fields calculated by TMM at the corresponding dip frequency $\omega_m$.
	GPs enter the GPC from $x$=0 where the area from $x=21d$ to $x=22.06d$ denotes the stopband. In Fig. \ref{FIG12}(b), the maximum enhancement of the electric field appears on the stopband and its enhancement factor is about 7.2.
	In Fig. \ref{FIG12}(c), the maximum enhancement of the magnetic field also appears on the stopband and its enhancement factor is similar to that of the electric field. Consistent with our expectations, when the GPs are at the lowest point of reflectance, the energy of GPs is trapped in the structure and the plasmonic Tamm state can be realized. The plasmonic Tamm state activated by GPC-stopband can achieve great concentration (the maximum enhancement of the electric field and the magnetic field all appear on the stopband) and get higher enhancement factor. Due to its high order tunability, GPC can provide a novel platform for the realization and enhancement of surface states such as plasmonic Tamm state, and TMM provides a new idea and simulation method for this novel optical phenomenon.  
	
	\section{Conclusion}
	We presented transfer matrix method (TMM) for graphene surface plasmonic crystals in this work. We found that, compared with the plane wave expansion method, TMM is more efficient in simulating certain photonic characteristics in the one-dimensional GPC structure. TMM is based on the local GPs, and takes into account the anomalous scattering current principle of jumping surface conductivity. We confirmed that the results obtained via TMM for the transmission and reflection spectra of the GPC are consistent with those obtained via plane wave expansion method. We also realized defect induced transparency due to the presence of a defect mode in the GPC. This makes the GPC highly transparent in specific frequency regions and can be tuned conveniently. In addition, we examined a specific example of the GPC platform to achieve the plasmonic Tamm state. The results show that GPC can be used as a novel platform for plasmonic Tamm state with small device size and high enhancement. Our analytical results and related physical insights into the plasmonic interaction with periodic surface conductivity will help the design of integrated absorbers and traps, providing references for micro-nano detection and surface fluorescence enhancement. \\
	
	\begin{acknowledgments}
		This work was supported in part by the National Natural Science
		Foundation of China under the Grant No. 11274132 and No. 11804219.\\
	\end{acknowledgments}
	
	\appendix
	\section{Reflection and transmission coefficients of surface current of the A-B structure }\label{Appendix A}
	Considering a one-dimensional case $ \vec{J}=J(x)\hat{x}$, the self-consistent equation for surface current is
	\begin{equation}\label{eqA1}
		\dfrac{J(x)}{\sigma(x)}=E^{ext}(x)+\dfrac{1}{4\pi\epsilon_0\epsilon_e}\iint^{\infty}_{-\infty}\dfrac{\rho(x')}{|\vec{r'}-\vec{r}|^3}(x'-x)dx'dy'.
	\end{equation}
	The distribution of $\sigma(x)$ is certain.  
	Using the conservation relationship $i\omega\rho(x)=\nabla \cdot \vec{J}$ and simplifying Eq. (\ref{eqA1}) to get
	\begin{equation}\label{eqA2}
		\dfrac{J(x)}{\sigma(x)}=E^{ext}(x)+\dfrac{1}{2\pi i \omega\epsilon_0\epsilon_e}\int^{\infty}_{-\infty}\dfrac{\partial_{x'} J(x')}{x'-x}dx' , 
	\end{equation}
	where $\epsilon_e=(\epsilon_1+\epsilon_2)/2$ represents the relative permittivity of the environment. The integral in Eq. (\ref{eqA2}) is of the Cauchy principal-value type. To deal with this kind of problem, we need to introduce one-sided Fourier transform  
	\begin{equation}\label{eqA4}
		\begin{split}
			F_L=&\int^0_{-\infty} \dfrac{J(x)}{2\pi} e^{-ikx} dx,\\ F_R=&-\int^{\infty}_0 \dfrac{J(x)}{2\pi} e^{-ikx} dx ,\\
			&F_L-F_R=J_k.
		\end{split}
	\end{equation}
	Since our purpose is to study the scattering of the surface current when it crosses the boundary of the Fermi energy mutation, we can simply assume that $\sigma(x<0)=\sigma_L$ and $\sigma(x>0)=\sigma_R$.
	Then, the Fourier transform of Eq. (\ref{eqA2}) can be written as 
	\begin{equation}\label{eqA3}
		F_L/\sigma_L -F_R/\sigma_R
		= E^{ext}_k + \dfrac{1}{i\omega\epsilon_0\epsilon_e} F\{\partial_xJ(x)\}  F\{\dfrac{1}{x}\}.
	\end{equation}
	
	Using the relations $F\{\partial_xJ(x)\}=ikJ_k(k)$ and $F\{\dfrac{1}{x}\}=-\dfrac{i}{2}sgn(k)$, Eq. (\ref{eqA3}) becomes
	\begin{equation}\label{eqA5}
		F_L/\sigma_L -F_R/\sigma_R = E^{ext}_k + \dfrac{|k|}{2i\omega\epsilon_0\epsilon_e} (F_L-F_R).
	\end{equation}
	Introducing the definitions $k_L=2i\omega\epsilon_0\epsilon_e/\sigma_L$ and  $k_R=2i\omega\epsilon_0\epsilon_e/\sigma_R$, Eq. (\ref{eqA5}) can be written in a concise form
	\begin{equation}\label{eqA6}
		F_L = \dfrac{k_R-|k|}{k_L-|k|}F_R+\dfrac{2i\omega\epsilon_o\epsilon_e}{k_L-|k|}E^{ext}_k=v(k)F_R+ s(k),
	\end{equation}
	where 
	\begin{equation}
		v(k)=\dfrac{k_R-|k|}{k_L-|k|}, \quad s(k)=\dfrac{2i\omega\epsilon_o\epsilon_e}{k_L-|k|}E^{ext}_k. 
	\end{equation}
	Note that if the conductivity of the entire plane is $\sigma_0$ without jumping, that is, $\sigma_L=\sigma_R=\sigma_0$, then the Fourier transform of the surface current can be easily obtained as 
	\begin{equation}\label{eqA7}
		J_{k}^0=\dfrac{2i\omega\epsilon_0\epsilon_e}{k_p-|k|}E^{ext}_{k} ,\quad
		k_p=\dfrac{2i\omega\epsilon_0\epsilon_e}{\sigma_0}.
	\end{equation}
	$J_k^0$ has a pole at $k=\pm k_p$, which implies that the eigen wave number of the wave in the absence of sources is $k_p$. 
	Comparing $J_k^0$ with $s(k)$, $s(k)$ can be understood as the intrinsic mode under the wave vector $k_L$.
	
	Now we suppose that the solution of Eq. (\ref{eqA6}) has the following form
	\begin{align}
		F_L(k)&=s(k)-\dfrac{X^+(k)}{2\pi i}\int\dfrac{s(q)dq}{(q-k)X^+(q)}\label{eqA8}\displaybreak\\
		F_R(k)&=-\dfrac{1}{v(k)}\dfrac{X^+(k)}{2\pi i}\int\dfrac{s(q)dq}{(q-k)X^+(q)}\label{eqA9}\\
		X^+(k)&=\dfrac{k_R+k}{k_L+k} exp\{-\dfrac{1}{4\pi i}\int \dfrac{ln[v(p)]dp}{p-k}\}\label{eqA10}
	\end{align}
	For Equation (\ref{eqA8}), $s(k)$ can be understood as the Fourier transform of the intrinsic current under the conductivity $\sigma_L$ for $x<0$. So the last term in Eq. (\ref{eqA8}) can be described as the Fourier transform of the reflected current.
	
	For the exponential term in Eq. (\ref{eqA10}), the following transformations can be made 
	\begin{widetext}
		\begin{align}\label{eqA12}
			-\dfrac{1}{4\pi i}\int \dfrac{ln[v(p)]dp}{p-k}
			&=i\Big\{\dfrac{1}{4\pi }\int \dfrac{ln[v(p)]dp}{p-k} \
			- \dfrac{i}{2}ln\Big( \dfrac{k_R+|k|}{k_L+|k|} \Big) \Big\}
			- \dfrac{1}{2}ln\Big( \dfrac{k_R+|k|}{k_L+|k|}\Big) \nonumber\\
			&=i\Big\{\dfrac{1}{4\pi }\int \dfrac{ln[v(p)]dp}{p-k} \
			- \dfrac{1}{4\pi}\int ln\Big( \dfrac{k_R+|p|}{k_L+|p|}\Big) \dfrac{dp}{p-k} \Big\}
			- \dfrac{1}{2}ln\Big( \dfrac{k_R+|k|}{k_L+|k|}\Big) \nonumber\\
			&=i\dfrac{1}{4\pi}\int ln\Big[ \dfrac{(k_R+|p|)(k_L-|p|)}{(k_L+|p|)(k_R-|p|)} \Big] \dfrac{dp}{k-p}
			- \dfrac{1}{2}ln\Big( \dfrac{k_R+|k|}{k_L+|k|}\Big) \nonumber\\
			&=i\dfrac{1}{4\pi}\Big\{\int_0^{\infty} ln\Big[ \dfrac{(k_R+|p|)(k_L-|p|)}{(k_L+|p|)(k_R-|p|)} \Big] \dfrac{dp}{k-p}
			+ \int^0_{-\infty} ln\Big[ \dfrac{(k_R+|p|)(k_L-|p|)}{(k_L+|p|)(k_R-|p|)} \Big] \dfrac{dp}{k-p} \Big\}- \dfrac{1}{2}ln\Big( \dfrac{k_R+|k|}{k_L+|k|}\Big) \nonumber\\
			&=i\dfrac{1}{4\pi}\int^{\infty}_0 ln\Big[ \dfrac{(k_R+|p|)(k_L-|p|)}{(k_L+|p|)(k_R-|p|)} \Big]
			(\dfrac{1}{p+k}+\dfrac{1}{k-p})dp
			- \dfrac{1}{2}ln\Big( \dfrac{k_R+|k|}{k_L+|k|}\Big) \nonumber\\
			&=-i\dfrac{k}{2\pi}\int^{\infty}_0 ln\Big[ \dfrac{(k_R+|p|)(k_L-|p|)}{(k_L+|p|)(k_R-|p|)} \Big]\dfrac{dp}{p^2-k^2}
			- \dfrac{1}{2}ln\Big( \dfrac{k_R+|k|}{k_L+|k|}\Big) \nonumber\\
			&=-i\dfrac{k}{2\pi}\int^{\infty}_0 \Big[ ln \Big(\dfrac{k_R+p}{k_L+p}\Big) -ln\Big( \dfrac{k_R-p}{k_L-p}\Big)\Big]\dfrac{dp}{p^2-k^2}
			- \dfrac{1}{2}ln\Big( \dfrac{k_R+|k|}{k_L+|k|}\Big) \nonumber\\
			&=i\Big[-\dfrac{k}{\pi}\int^{\infty}_0ln\Big( \dfrac{k_R+p}{k_L+p} \Big) \dfrac{dp}{p^2-k^2}
			+\dfrac{k}{2\pi}\int^{\infty}_0ln\Big( \dfrac{k_R+p}{k_L+p} \Big) \dfrac{dp}{p^2-k^2}\nonumber\\
			&\quad\quad\quad\quad\quad\quad\quad\quad\quad\quad\quad\quad+\dfrac{k}{2\pi}\int^{\infty}_0ln\Big( \dfrac{k_R-p}{k_L-p} \Big) \dfrac{dp}{p^2-k^2}\Big]
			- \dfrac{1}{2}ln\Big( \dfrac{k_R+|k|}{k_L+|k|}\Big) \\
			&=i\Big[-\dfrac{k}{\pi}\int^{\infty}_0ln\Big( \dfrac{k_R+p}{k_L+p} \Big) \dfrac{dp}{p^2-k^2}
			-\dfrac{k}{2\pi}\int^{\infty}_0ln\Big( \dfrac{k_R^2-p^2}{k_L^2-p^2}\Big)\dfrac{dp}{k^2-p^2}\Big]
			- \dfrac{1}{2}ln\Big( \dfrac{k_R+|k|}{k_L+|k|}\Big) \nonumber\\
			&=i\Big[-\dfrac{k}{\pi}\int^{\infty}_0ln\Big( \dfrac{k_R+p}{k_L+p} \Big) \dfrac{dp}{p^2-k^2}
			-\dfrac{k}{2\pi i}\int^{i\infty}_0ln\Big( \dfrac{k_R^2+p^2}{k_L^2+p^2}\Big)\dfrac{dp}{k^2+p^2}\Big]
			- \dfrac{1}{2}ln\Big( \dfrac{k_R+|k|}{k_L+|k|}\Big) \nonumber\\
			&=i\Big[-\dfrac{k}{\pi}\int^{\infty}_0ln\Big( \dfrac{k_R+p}{k_L+p} \Big) \dfrac{dp}{p^2-k^2}
			-\dfrac{k}{2\pi i}\int^{\infty}_0ln\Big( \dfrac{k_R^2+p^2}{k_L^2+p^2}\Big)\dfrac{dp}{k^2+p^2}\Big]
			- \dfrac{1}{2}ln\Big( \dfrac{k_R+|k|}{k_L+|k|}\Big) \nonumber\\
			&=i\Big[-\dfrac{k}{\pi}\int^{-i\infty}_0ln\Big( \dfrac{k_R+p}{k_L+p} \Big) \dfrac{dp}{p^2-k^2}
			-\dfrac{k}{2\pi i}\int^{\infty}_0ln\Big( \dfrac{k_R^2+p^2}{k_L^2+p^2}\Big)\dfrac{dp}{k^2+p^2}\Big]
			- \dfrac{1}{2}ln\Big( \dfrac{k_R+|k|}{k_L+|k|}\Big) \nonumber\\
			&=i\Big[-\dfrac{ik}{\pi}\int^{\infty}_0ln\Big( \dfrac{k_R-ip}{k_L-ip} \Big) \dfrac{dp}{p^2+k^2}
			-\dfrac{k}{2\pi i}\int^{\infty}_0ln\Big( \dfrac{k_R^2+p^2}{k_L^2+p^2}\Big)\dfrac{dp}{k^2+p^2}\Big]
			- \dfrac{1}{2}ln\Big( \dfrac{k_R+|k|}{k_L+|k|}\Big) \nonumber\\
			&=i\dfrac{k}{\pi}\int^{\infty}_0\Big\{ \dfrac{1}{2} ln\Big( \dfrac{k_R^2+p^2}{k_L^2+p^2} \Big)+i[arctan(\dfrac{p}{k_R})-arctan(\dfrac{p}{k_L})]\Big\} \dfrac{dp}{p^2+k^2}\nonumber
			\\&\quad\quad\quad\quad\quad\quad\quad\quad\quad\quad\quad\quad-\dfrac{k}{2\pi}\int^{\infty}_0ln\Big( \dfrac{k_R^2+p^2}{k_L^2+p^2}\Big)\dfrac{dp}{k^2+p^2}
			-  \dfrac{1}{2}ln\Big( \dfrac{k_R+|k|}{k_L+|k|}\Big) \nonumber\\
			&=\dfrac{k}{\pi}\int^{\infty}_0arctan[\dfrac{(k_L-k_R)p}{k_Lk_R+p^2}]\dfrac{dp}{p^2+k^2}
			-\dfrac{1}{2}ln\Big( \dfrac{k_R+|k|}{k_L+|k|}\Big)\nonumber
			\\
			&=i\phi(k)-\dfrac{1}{2}ln\Big( \dfrac{k_R+|k|}{k_L+|k|}\Big)\nonumber
		\end{align}
	\end{widetext}
	where we assume
	\begin{equation}\label{eqA13}
		\phi(k)=\dfrac{k}{\pi}\int^{\infty}_0 arctan[\dfrac{(k_L-k_R)p}{k_Lk_R+p^2}]\dfrac{dp}{p^2+k^2},
	\end{equation}
	and $k_L<k_R$.
	Inserting Eq. (\ref{eqA12}) into Eq. (\ref{eqA10}) to get
	\begin{align}\label{eqA14}
		\begin{split}
			X^+(k)= e^{i\phi(k)} \Big( \dfrac{k_L+|k|}{k_R+|k|} \Big)^{1/2} \dfrac{k_R+k}{k_L+k} .
		\end{split}
	\end{align}
	In order to obtain the reflection and transmission properties of the system, we assume an incident current
	\begin{equation}\label{eqA15}
		J^{i,L}(x)=2\pi Ae^{ik_Lx},
	\end{equation}
	with Fourier transform
	\begin{equation}\label{eqA16}
		s(k)=2\pi A\delta(k_L-k).
	\end{equation}
	Inserting it into the last item in (\ref{eqA8})
	\begin{equation}\label{eqA17}
		\begin{split}
			F_L(k)=&s(k)-\dfrac{1}{2\pi i}  \int\dfrac{X^+(k)s(q)dq}{(q-k)X^+(q)}\\
			=&s(k)+J^{s,L}_k,
		\end{split}
	\end{equation}
	where
	\begin{equation}\label{eqA18}
		\begin{split}
			J_k^{s,L}
			&=-\dfrac{1}{2\pi i}\int\dfrac{X^+(k)s(q)dq}{(q-k)X^+(q)}\\
			&=-\dfrac{iA}{X^+(k_L)}\dfrac{X^+(k)}{k-k_L}\\
			&=-\dfrac{iAX^+(k)}{(k-k_L)X^+(k_L)}.
		\end{split}
	\end{equation}
	
	The inverse Fourier transform of Eq. (\ref{eqA18}) is
	\begin{equation}\label{eqA19}
		\begin{split}
			J^{s,L}(x)&=\int J_k^{s,L} e^{ikx} dk \\
			&= - \int \dfrac{iAX^+(k)e^{ikx}}{(k-k_L)X^+(k_L)}dk  \\
			&= -\int \dfrac{iAe^{i\phi(k)}(k_L+|k|)^{1/2}(k_R+k)e^{ikx}}{(k_R+|k|)^{1/2}X^+(k_L)(k-k_L)(k+k_L)}dk\\
			&=-2\pi i\dfrac{iAe^{i\phi(-k_L)}(k_L+|k_L|)^{1/2}(k_R-k_L)e^{-ik_Lx}}{(k_R+|k_L|)^{1/2}X^+(k_L)(-2k_L)}\\
			&=-\dfrac{\pi Ae^{i\phi(-k_L)}(k_L+|k_L|)^{1/2}(k_R-k_L)e^{-ik_Lx}}{(k_R+|k_L|)^{1/2}k_LX^+(k_L)}\\
			&=2\pi A e^{2i\phi(-k_L)}\dfrac{k_L-k_R}{k_R+k_L}e^{-ik_Lx}.
		\end{split}
	\end{equation}
	
	Comparing with Eq. (\ref{eqA15}), it can be found that the reflection coefficient of the incident current is  $r=e^{i2\phi(-k_L)}\dfrac{k_L-k_R}{k_R+k_L}$. According to the definition of $\phi(k)$, we can simplify the phase in $r$
	\begin{equation}\label{eqA20}
		\begin{split}
			2\phi(-k_L)&=-\dfrac{2k_L}{\pi}\int^{\infty}_0 arctan[\dfrac{(k_L-k_R)p}{k_Rk_L+p^2}]\dfrac{dp}{p^2+k_L^2}\\
			&=-\dfrac{2}{\pi}\int^{\infty}_0arctan[\dfrac{pk_L/k_R-p}{1+p^2k_L/k_R}]\dfrac{dp}{p^2+1}\\
			&=-\dfrac{2}{\pi}\int^{\infty}_0[arctan(pk_L/k_R)-arctan(p)]\dfrac{dp}{p^2+1}\\
			&=\dfrac{2}{\pi}\int^{\infty}_0\dfrac{arctan(p)}{p^2+1}dp
			-\dfrac{2}{\pi}\int^{\infty}_0\dfrac{arctan(pk_L/k_R)}{p^2+1}dp\\
			&=\dfrac{\pi}{4}-\dfrac{2}{\pi}\int^{\infty}_0\dfrac{arctan(pk_L/k_R)}{p^2+1}dp\\
			&=\theta
		\end{split}
	\end{equation}
	For $x>0$ area
	\begin{equation}\label{eqA21}
		J^{t,R}_k=-F_R=\dfrac{iA}{X^+(k_L)v(k)}\dfrac{X^+(k)}{k-k_L},
	\end{equation}
	its inverse Fourier transform is 
	\begin{equation}\label{eqA22}
		\begin{split}
			J^{t,R}(x)&=\int J^{t,R}_k e^{ikx} dk\\
			&=\int \dfrac{iAX^+(k)}{X^+(k_L)(k-k_L)}\dfrac{k_L-|k|}{k_R-|k|}e^{ikx}dk\\
			&=\int \dfrac{iAe^{i\phi(k)}(\dfrac{k_L+|k|}{k_R+|k|})^{1/2}(k_R+k)}{X^+(k_L)(k-k_L)(k+k_L)}\dfrac{k_L-|k|}{k_R-|k|}e^{ikx}dk.
		\end{split}
	\end{equation}
	For the integral in Eq. (\ref{eqA22}), we think that the result is mainly contributed by the pole at $k=k_R$. Therefore, Eq. (\ref{eqA22}) can be obtained from the residue and the result is
	\begin{equation}\label{eqA23}
		\begin{split}
			J^{t,R}(x)&=\int \dfrac{iAX^+(k)}{X^+(k_L)(k-k_L)}\dfrac{k_L-|k|}{k_R-|k|}e^{ikx}dk\\
			&=2\pi A \dfrac{\sqrt{2k_Lk_R}}{k_L+k_R}e^{ik_Rx}e^{i[\phi(k_R)-\phi(k_L)]}.
		\end{split}
	\end{equation}
	The relationship $\phi(k_R)=\phi(k_L)$ can be easily verified by the calculation similar to Eq. (\ref{eqA20}). So we can get the current expression at the $x>0$ area as
	\begin{equation}\label{eqA24}
		J^{t,R}(x)=2\pi A \dfrac{\sqrt{2k_Lk_R}}{k_L+k_R}e^{ik_R x},
	\end{equation}
	and the transmission coefficient of the incident current as
	\begin{equation}
		t=\dfrac{\sqrt{2k_Lk_R}}{k_L+k_R}.
	\end{equation}

	\section{Plane wave expansion method}\label{appendixB}
	First, the self-consistent equations describing the plasmonic excitation is introduced here~\cite{RN33}
	\begin{equation}\label{eqB1}
		\begin{aligned}
			\phi_{tot}(\omega,\vec{k})=&\phi_{ext}(\omega,\vec{k})+\sum_{\vec{k}'}\tilde{v}(\omega,\vec{k},\vec{k'})n(\omega,\vec{k}),\\
			n(\omega,\vec{k}&)=\sum_{\vec{k}'}\tilde{\chi}(\omega,\vec{k},\vec{k}')\phi_{tot}(\omega,\vec{k}'),
		\end{aligned}
	\end{equation}
	where $\tilde{v}(\omega,\vec{k},\vec{k'})$ is the effective screened Coulomb interaction, $\tilde{\chi}(\omega,\vec{k},\vec{k}')$ is the density response function of graphene and $\phi_{ext}(\omega,\vec{k})$ is the external driving potential~\cite{PhysRevB.96.035433}. 
	\begin{figure}[ht]
		\centering
		\includegraphics[width=\linewidth]{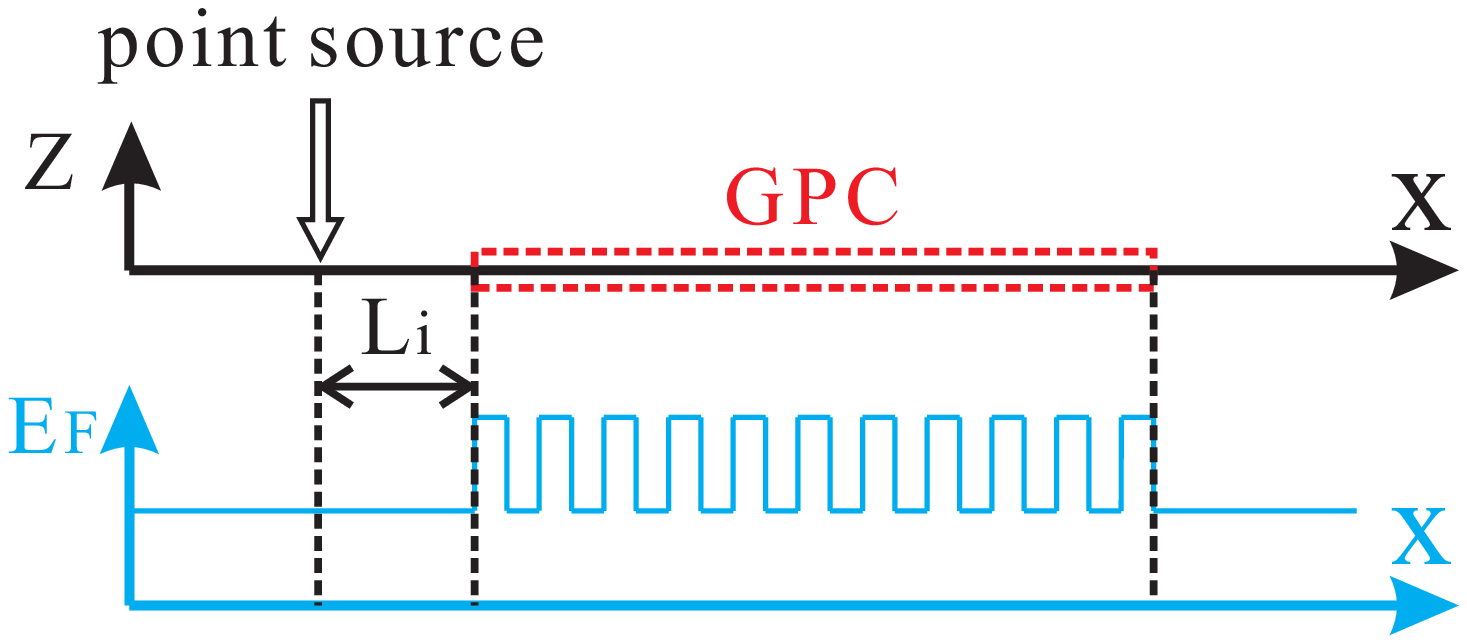}\\
		\begin{flushleft}
			FIG. B1. Graphene on the entire x-y plane, periodic point sources with a big period $D$. There is a GPC  at a distance of $L_i$ to the right of each point source. The blue coordinates show the Fermi energy distribution at the corresponding position of graphene. 
		\end{flushleft}
		\label{FIGB1}
	\end{figure}\\
	As shown in Fig. \ref{FIGB1}1, a periodic point source is added with a period length of D. In order to prevent modes other than GPs excited by it from entering the right side of the GPC, the point source is a certain distance away from the GPC on the right. The point source is far away from the GPC on the left in order to prevent its excitation modes from interfering with the left GPC, and the excitation mode will naturally dissipate during propagation.
	
	In this way, the field distribution of the point source excitation field transmitted into the GPC can be simulated by plane wave expansion method. For one-dimensional Bloch state with Bloch wave vector $\textbf{k}$, the momentum integral becomes a summation over a countable set, $\textbf{k}+\{ \textbf{G}_D \}$ ( $\{ \textbf{G}_D \}$ is the set of the reciprocal lattice vector corresponding to the large period $D$). Then, Eq. (\ref{eqB1}) are casted into an eigenvalue problem on discrete matrices~\cite{PhysRevLett.121.086807}
	\begin{equation}\label{eqB4}
		\begin{aligned}
			\widehat{V}_\textbf{k}\widehat{X}_\textbf{k}|\phi^{tot}&_\textbf{k}\rangle=\omega(\omega+i\tau^{-1})(|\phi^{tot}_\textbf{k}\rangle-|\phi^{ext}_\textbf{k}\rangle),\\
			[\widehat{V}&_\textbf{k}]_{\alpha,\beta}=\dfrac{e^2}{2\epsilon_e\epsilon_0|\textbf{k}+\textbf{G}_{D}^{\alpha}|\delta_{\alpha\beta}},\\
			[\widehat{X}_\textbf{k}&]_{\alpha,\beta}=\dfrac{\textbf{k}_{\alpha}\cdot\textbf{k}_{\beta}\tilde{E}_{F}(\textbf{k}_{\alpha}-\textbf{k}_{\beta})}{\pi\hbar^2},
		\end{aligned}
	\end{equation}
	where
	\begin{equation}\label{eqB3}
		\begin{aligned}
			\textbf{k}_{\alpha,\beta}&=\textbf{k}+\textbf{G}^{\alpha,\beta}_D,\\
			\tilde{E}_{F}(\textbf{k})=\dfrac{1}{D}&\int_{D}E_F(\textbf{r})e^{-i\textbf{k}\cdot\textbf{r}}d\textbf{r}.
		\end{aligned}
	\end{equation}
	Since the external field of the point source is periodic, $|\phi^{ext}_\textbf{k}\rangle$ can be easily obtained by Fourier integration, so that $|\phi^{tot}_\textbf{k}\rangle$ can be solved by using (\ref{eqB2}), and then $\phi_{tot}$ can be obtained by Fourier expansion
	\begin{equation}\label{eqB2}
		\phi_{tot}(\textbf{r})=\sum_{\alpha}e^{i\textbf{k}_\alpha\cdot\textbf{r}}[|\phi^{tot}_\textbf{k}\rangle]_\alpha.
	\end{equation}
	
	\nocite{*}
	
	\bibliography{xxx}
	
\end{document}